\documentclass{article}

\usepackage{PRIMEarxiv}
\usepackage[utf8]{inputenc} 
\usepackage[T1]{fontenc}    
\usepackage{hyperref}       
\usepackage{url}            
\usepackage{booktabs}       
\usepackage{nicefrac}       
\usepackage{microtype}      
\usepackage{fancyhdr}       
\usepackage{graphicx}       
\graphicspath{{media/}}     
\usepackage{amsmath,amsfonts}
\usepackage{algorithmic}
\usepackage{algorithm}
\usepackage{array}
\usepackage[caption=false,font=footnotesize]{subfig}  
\usepackage{textcomp}
\usepackage{stfloats}
\usepackage{url}
\usepackage{cite}
\usepackage{tabularx}      
\usepackage{caption}       
\usepackage{multirow}

\title{Beyond Identity: Generalizable Deepfake Audio Detection}

\author{
\begin{minipage}[t]{0.3\linewidth}
\centering
\textbf{Yasaman Ahmadiadli\textsuperscript{1*}}\\
\texttt{yahmadiadli@torontomu.ca}
\end{minipage}
\hfill
\begin{minipage}[t]{0.3\linewidth}
\centering
\textbf{Xiao-Ping Zhang\textsuperscript{1}}\\
\texttt{xpzhang@ieee.org}
\end{minipage}
\hfill
\begin{minipage}[t]{0.3\linewidth}
\centering
\textbf{Naimul Khan\textsuperscript{1}} \\
\texttt{n77khan@torontomu.ca}
\end{minipage}
}

\begin{document}

\maketitle
\footnotetext[1]{Department of Electrical, Computer and Biomedical Engineering, Toronto Metropolitan University, Toronto, Canada.}
\footnotetext[2]{*Corresponding author.}

\begin{abstract}
Deepfake audio presents a growing threat to digital security, due to its potential for social engineering, fraud, and identity misuse. However, existing detection models suffer from poor generalization across datasets, due to implicit identity leakage, where models inadvertently learn speaker-specific features instead of manipulation artifacts. To the best of our knowledge, this is the first study to explicitly analyze and address identity leakage in the audio deepfake detection domain. This work proposes an identity-independent audio deepfake detection framework that mitigates identity leakage by encouraging the model to focus on forgery-specific artifacts instead of overfitting to speaker traits. Our approach leverages Artifact Detection Modules (ADMs) to isolate synthetic artifacts in both time and frequency domains, enhancing cross-dataset generalization. We introduce novel dynamic artifact generation techniques, including frequency domain swaps, time domain manipulations, and background noise augmentation, to enforce learning of dataset-invariant features. Extensive experiments conducted on ASVspoof2019, ADD 2022, FoR, and In-The-Wild datasets demonstrate that the proposed ADM-enhanced models achieve F1 scores of 0.230 (ADD 2022), 0.604 (FoR), and 0.813 (In-The-Wild), consistently outperforming the baseline. Dynamic Frequency Swap proves to be the most effective strategy across diverse conditions. These findings emphasize the value of artifact-based learning in mitigating implicit identity leakage for more generalizable audio deepfake detection.
\end{abstract}

\keywords{Audio Deepfake Detection,\ Identity-Independent Detection,\ Artifact Detection,\ Cross-Dataset Generalization,\ Speaker Identity}

\section{Introduction}
\label{introduction}

Deepfake audio has emerged as a concerning threat, targeting voice-enabled authentication systems, financial fraud, and social engineering attacks \cite{shahzad2022review}. Leveraging advanced generative models, these synthetic speech samples achieve near-human realism, making detection increasingly challenging. The ability of deepfake audio to impersonate individuals with high fidelity introduces new vulnerabilities across security-sensitive domains \cite{liz2024generation, shaaban2023audio}.

While numerous detection methods have been proposed to distinguish between real and fake media, many suffer from significant performance degradation in cross-dataset evaluations. Existing approaches primarily focus on architectural improvements \cite{tak2021end, Chen2021urchannel, tak2021endg, AASIST} or training process optimizations such as data augmentation \cite{Park2019Specaugment,rawboost}. Despite these advances, deepfake detectors often fail to generalize in real-world scenarios. This is primarily due to a phenomenon known as implicit identity leakage, where models inadvertently rely on speaker-specific characteristics rather than manipulation artifacts \cite{dong2023implicit}.

As a result, these models excel at detecting deepfakes within the same dataset (intra-dataset performance) but perform poorly when tested on unseen speakers or new generative models (cross-dataset evaluation) \cite{tinsley2021face}. This issue is particularly severe for non-public figures, whose voices may not be included in training datasets, leading to severely degraded detection performance in real-world applications.

In deepfake detection, feature extraction plays a fundamental role in distinguishing real from fake media. For video and image-based deepfake detection, models typically rely on facial cues such as eye blinking, head pose, and facial expressions \cite{survey}. In contrast, audio deepfake detection depends on spectral and cepstral features such as Mel-Frequency Cepstral Coefficients (MFCCs), Mel-spectrograms, pitch contours, and Constant-Q Cepstral Coefficients (CQCCs) \cite{li2022comparative}. However, these features do not explicitly address identity leakage, reinforcing the need for identity-independent detection strategies that prioritize artifact-based learning over speaker-specific cues.

To address the challenge of identity leakage in audio deepfake detection, we introduce an identity-independent detection framework. Inspired by the findings in \cite{dong2023implicit}, which confirmed identity leakage in video deepfake detection, we systematically investigate this phenomenon in the audio domain. We propose a novel artifact-based detection scheme inspired by prior work on visual deepfake identity leakage. Our approach leverages \textit{Artifact Detection Modules} (ADMs), designed to isolate synthetic artifacts rather than identity-specific traits, improving cross-dataset generalization.

The ADM framework introduces artifact-centric learning, where synthetic artifacts are generated and analyzed in both frequency and time domains (e.g., frequency band swaps between real and fake samples of the same speaker, and background noise perturbations). These artifacts act as forgery markers, allowing the model to distinguish fake audio based on structural inconsistencies rather than identity-related features. Although speaker-matched artifact generation may appear counterintuitive for identity-independent detection, this design ensures that the only artificial variation introduced is due to controlled artifacts, guiding the model to detect manipulations while holding speaker identity constant during artifact creation. This design ensures that the model attends to forgery-related inconsistencies without being biased by speaker identity, as motivated by prior findings in the visual deepfake domain. The primary contributions of this work are as follows:

\begin{itemize}
    \item \textbf{First Analysis of Identity Leakage in Audio Deepfake Detection:} 
    This is the first work to \textit{explicitly} analyze and tackle identity leakage within the \textit{audio-based} deepfake detection domain, bridging a gap in the literature where most prior work focused on face or image-based deepfakes.
    \item \textbf{Artifact Detection Modules (ADMs):} 
    We introduce novel artifact-centric learning modules, specifically designed to isolate and amplify forgery cues (instead of speaker identity). These modules integrate time and frequency domain manipulations for robust artifact generation.
    \item \textbf{Dynamic Artifact Generation:}
    We propose advanced augmentation techniques (e.g., dynamic frequency swaps, background noise injection) that enforce dataset-invariant feature learning, helping the model to sidestep identity cues.
    \item \textbf{Improved Cross-Dataset Generalization:}
    Experimental results across four benchmark datasets (ASVspoof2019, ADD 2022, FoR, and In-The-Wild) confirm that the proposed approach achieves superior generalization compared to baseline methods. 
\end{itemize}

\section{Related Work}
\label{sec:related}

This section provides an overview of significant advances in audio deepfake detection, focusing on cross-dataset generalization and the persistent challenge of identity leakage. While recent methods often succeed in capturing subtle artifacts left by generative models, many approaches still rely heavily on speaker-dependent cues, such as voice identity or channel characteristics, to achieve high accuracy. This reliance can compromise generalization when encountering new speakers or unseen attack types. Although some works have proposed partial solutions, these typically require large amounts of annotated data or still encode residual identity traces. Consequently, truly identity-agnostic detection pipelines remain an underexplored research direction, highlighting the need for more robust strategies that isolate manipulation artifacts from speaker-specific information.

Audio deepfake detection has progressed from early approaches using hand-crafted features like MFCCs and CQCCs \cite{kinnunen2020tandem, sahidullah15_interspeech, surveyACM, todisco2017constant} to deep learning-based models designed to detect artifacts produced by high-fidelity synthesis methods such as WaveNet and Tacotron \cite{van2016wavenet, wang17n_interspeech}. Although CNN-based architectures using mel-spectrograms \cite{fathan2022mel} and augmentation \cite{wang2020asvspoof} have improved robustness under channel mismatch, many still exhibit poor cross-dataset generalization \cite{yi2022add}.

Xception-based models have become popular due to their fine-grained spectral feature extraction using depthwise-separable convolutions \cite{muller22_interspeech, yi2022add}. However, as highlighted in \cite{liu2023asvspoof}, such models are prone to overfitting, often relying on speaker-specific or channel-dependent cues rather than universal spoofing artifacts. More adaptive architectures like neural stitching \cite{neuralstitch} combine shallow and deep layers to capture both generic and task-specific features, but still depend on dataset-specific preprocessing pipelines, limiting scalability.

A key but underexplored challenge is \textit{identity leakage}, where detection models inadvertently encode speaker identity (e.g., timbre, prosody) rather than real spoof artifacts \cite{khanjani2021deep}. Some methods attempt to mitigate this via adversarial training or multi-task learning \cite{tak2021end, mcuba2023effect}, but their reliance on large labeled datasets limits scalability and adaptability to unseen speakers. In the visual domain, \cite{dong2023implicit} recently demonstrated that identity cues overshadow manipulation artifacts in video-based deepfake detection. This motivates our exclusive focus on identity leakage in audio-based detection systems, which remains underexplored compared to visual domains.

We propose a system to tackle the identity leakage issue in audio deepfake detection, designing an Artifact Detection Module (ADM) that targets local synthetic distortions over identity traits. This approach is similar to \cite{cozzolino2023audio, cozzolino2021id}, which analyzed identity awareness in multimodal or face-based deepfake detection, but differs fundamentally by isolating audio artifacts (e.g., frequency and time manipulations) to encourage identity-independent detection. As a result, our proposed framework better handles unseen speakers and emergent generative methods.

\section{Proposed Method}
\label{sec:proposed}

\subsection{Problem Definition} \label{sec:problem_definition}

This work addresses implicit identity leakage in audio deepfake detection by designing an identity-independent detection system. The task is formulated as a binary classification problem over datasets \( X \) and \( Y \), where: \\
\( X = \{x_i | i = 1, \dots, N\}\): a set of real samples.\\
\( Y = \{y_j | j = 1, \dots, M\}\): a set of fake samples.\\
\( Z = \{z_k | k = 1, \dots, K\}\): a set of artifact-fake samples generated by augmenting \( Y \) with artifact information.

The goal is to train a detector \( D \) that minimizes dependency on subject-specific features within \( X \) and \( Y \), instead focusing on dataset-invariant artifact information within \( Z \). This encourages the model to learn forgery cues that generalize across speakers and datasets. We propose an Artifact Detection Module (ADM) that learns to focus on synthetic inconsistencies, improving cross-dataset generalizability. To achieve this, we define two training objectives:
\textit{ADM training to distinguish original fakes \(Y\) from artifact fakes \(Z\)}:

\begin{equation}
\min_{\theta_{\text{ADM}}} \mathbb{E}_{y \sim Y, z \sim Z} \left[ L(D_{\theta_{\text{ADM}}}(y), 0) + L(D_{\theta_{\text{ADM}}}(z), 1) \right]
\end{equation}
\textit{Main detector training to separate real \(X\) and fake \(Y\)}:

\begin{equation}
\min_{\theta} \mathbb{E}_{x \sim X, y \sim Y} \left[ L(D_{\theta}(x), 1) + L(D_{\theta}(y), 0) \right]
\end{equation}

where \( \theta \) represents the model parameters that are optimized.\\ \( \mathbb{E} \) denotes the expectation operator, which averages over samples from the datasets.\\
\( x \sim X \), \( y \sim Y \), \( z \sim Z \) are random samples drawn from the real dataset \( X \), fake dataset \( Y \), and artifact-fake dataset \( Z \), respectively.\\
\( L \) is the Binary Cross-Entropy Loss function, with:
\begin{itemize}
    \item \( L(D_{\theta}(x), 1) \): Evaluating the detector \( D \) when a real sample \( x \) is correctly classified as real (1).
    \item \( L(D_{\theta}(y), 0) \): Evaluating the detector \( D \) when a fake sample \( y \) is correctly classified as fake (0).
    \item \( L(D_{\theta}(z), 1) \): Evaluating the detector \( D \) when an artifact-enhanced fake sample \( z \) is classified as artifact-fake (1).
\end{itemize} 

This two-stage learning encourages the model to learn dataset-invariant features from artifact-rich regions while minimizing identity leakage, supporting cross-dataset generalizability. Our analysis focuses solely on audio, evaluated across ASVspoof2019 LA, ADD 2022, FoR, and In-The-Wild.

\subsection{Overview of the Audio Deepfake Detection Scheme}

\begin{figure*} [htbp]
\centering
\includegraphics[width=\textwidth]{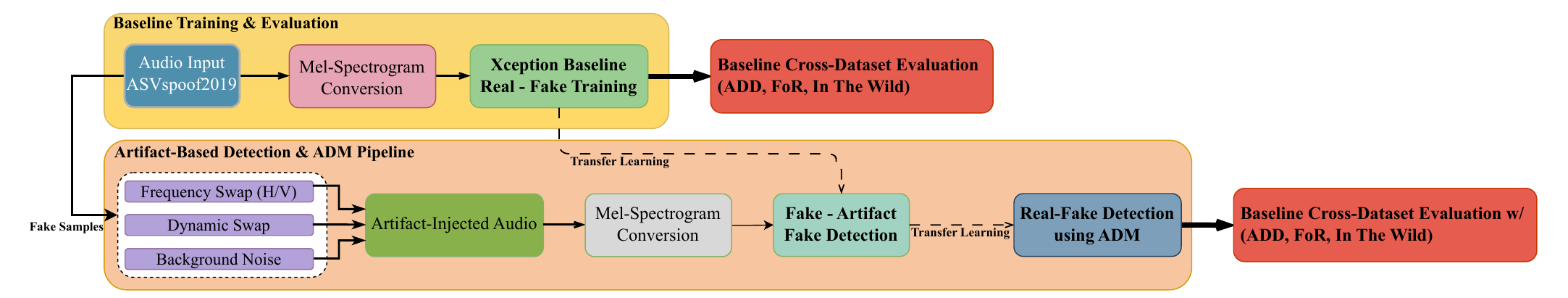}
\caption{Overview of the proposed pipeline. The process begins with standardized audio input, which is converted into Mel-spectrogram images for training an Xception-based baseline model. This baseline is evaluated on multiple datasets to assess its generalization ability. In the second stage, artifacts are introduced into fake audio samples using frequency-based swaps (horizontal and vertical), dynamic frequency swaps, and background noise addition. The Artifact Detection Module (ADM) is trained to differentiate between fake and artifact-injected samples. This module is then fine-tuned for real-fake detection using artifact-informed representations. Finally, cross-dataset evaluations are conducted using the ADM-infused model to measure its effectiveness.} 

\label{fig2}
\end{figure*}

The proposed scheme includes the following main components:\\
\textbf{Artifact Detection Module (ADM):} This module is trained to detect synthetic artifacts in deepfake audio. By emphasizing artifacts rather than identity cues, it supports improved cross-dataset performance. The artifact generation is performed between real and fake samples of the same speaker to mitigate identity dependence, ensuring that models generalize to unseen voices rather than learning speaker-specific patterns.\\
\textbf{Artifact Generation Techniques:} Introduce diverse, localized perturbations including frequency swaps, time-domain segment replacements, and background noise between fake and real samples of the same speaker to maintain identity-independence. These techniques create subtle artifacts that highlight inconsistencies in deepfake audio, making detection more robust against unseen attacks.\\
\textbf{Cross-Dataset Evaluation:} The proposed framework is trained on ASVspoof2019 LA and first evaluated on the same dataset to establish intra-dataset performance. To assess its generalizability, the model is then fine-tuned and tested on ADD 2022, FoR, and In-The-Wild datasets.This evaluation assesses whether ADM-based learning effectively reduces identity leakage and improves generalization. The overall architecture is illustrated in Fig~\ref{fig2} and later discussed in relation to generalization performance in Section~\ref{sec:results}.

\captionsetup{font=small}

\subsection{Audio Artifact Detection Module (ADM)}

The ADM shares the same architecture as the baseline Xception model but is trained to classify between original fake samples and their artifact-enhanced versions. The Xception model’s ability to capture spatial feature representations across channels enhances the detection performance of ADM. These artifacts simulate subtle synthetic manipulations without altering speaker identity. The learned representations are then transferred to the main fake-vs-real detector to improve robustness.

\subsubsection{Artifact Generation Techniques} 
\begin{figure*}[htbp]
  \centering

  \begin{minipage}[b]{0.48\linewidth}
    \centering
    \subfloat[Real sample\label{fig:genuine}]{
      \includegraphics[height=3cm]{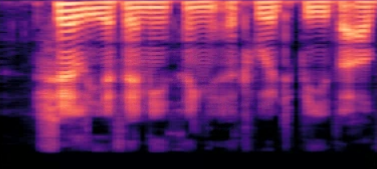}
    }
  \end{minipage}
  \hfill
  \begin{minipage}[b]{0.48\linewidth}
    \centering
    \subfloat[Fake sample\label{fig:fake}]{
      \includegraphics[height=3cm]{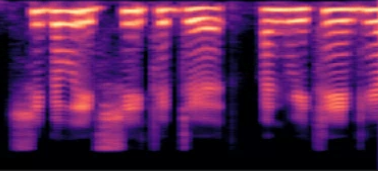}
    }
  \end{minipage}

  \caption{Mel-spectrogram examples from ASVspoof2019. Dark purple indicates low intensity; bright yellow, high energy.}
  \label{fig:mel_spectrograms}
\end{figure*}

As illustrated in Fig. \ref{fig:mel_spectrograms}, artifacts in the Mel-spectrogram representations of fake speech are not visually apparent. To address this, we manually insert artifacts in both the frequency and time domains, generating artifact-augmented fake samples (\( Z \)) to train the ADM. The artifact insertion process ensures that the fake sample remains perceptually realistic, preventing it from being transformed into an entirely different sample. Artifacts are generated by modifying fake samples using real ones from the same speaker ID in ASVspoof2019 dataset, ensuring identity-neutral transformations and avoiding cross-speaker variations. We apply three audio-domain augmentation strategies: fixed and dynamic frequency swaps, time-domain segment replacements, and background noise additions. Each method introduces perturbations that preserve speaker identity while revealing spoofing irregularities. This artifact-aware representation is then used to fine-tune a deepfake detector focused on spoofing cues rather than speaker traits.

\subsubsection{Fixed Frequency Swap Artifacts}

Since Mel-spectrograms are visual representations of audio signals, with the Y-axis representing frequency and the X-axis representing time, modifying the frequency domain introduces horizontal perturbations that highlight spectral inconsistencies.

There is no one-to-one correspondence between  real and fake samples in ASVspoof2019. To create frequency-domain artifacts, we randomly select a  real sample from the same speaker ID as the fake sample's ID and swap a specific frequency range between the two. The selected frequency bands correspond to key speech formants, ensuring realistic yet identifiable modifications.

Each speech sample contains Fundamental frequency (\( F_0 \)) the lowest frequency, First formant (\( F_1 \)) correlates with vowel openness, Second formant (\( F_2 \)) reflects tongue positioning along the front-back axis, Third formant (\( F_3 \)) associated with lip rounding and resonance.

In this work, we swap frequencies within the \( F_2 \) and \( F_3 \) ranges (2000-3500 Hz), as these formants capture significant phonetic variations \cite{cuccovillo2024audio}. The frequency swapping process is implemented as follows:

\begin{enumerate}
    \item \textbf{Standardization:} Both  real and fake samples are truncated or padded to 3 seconds for consistency.
    \item \textbf{FFT Transformation:} The Fast Fourier Transform (FFT) is applied to convert the time-domain signals into their frequency-domain representations:
\end{enumerate}

\begin{equation}
S(\omega) = \text{FFT}(s(t)), \quad B(\omega) = \text{FFT}(b(t)) \label{eq:fft}
\end{equation}

where \( s(t) \) and \( b(t) \) represent the fake and  real audio signals in the time domain, respectively.

\begin{enumerate}
    \setcounter{enumi}{2}
    \item \textbf{Frequency Bin Calculation:} The frequency bins corresponding to each FFT component are computed as:
\end{enumerate}

\begin{equation}
f_k = \frac{k \cdot SR}{N} \label{eq:freq_bins}
\end{equation}

where \( k \) is the FFT index, \( SR \) is the sampling rate, and \( N \) is the total number of samples.

\begin{enumerate}
    \setcounter{enumi}{3}
    \item \textbf{Selecting Frequency Band for Swapping:} The start and end indices of the targeted frequency range are determined using:
\end{enumerate}

\begin{equation}
\text{start\_index} = \min \left\{ k : f_k \geq F_{\text{start}} \right\} \label{eq:start_index}
\end{equation}

\begin{equation}
\text{end\_index} = \min \left\{ k : f_k \geq F_{\text{end}} \right\} \label{eq:end_index}
\end{equation}

where \( F_{\text{start}} \) and \( F_{\text{end}} \) define the frequency range.

\begin{enumerate}
    \setcounter{enumi}{4}
    \item \textbf{Replacing Fake Sample Frequencies:} The FFT components of the  real sample replace those in the fake sample for the selected range:
\end{enumerate}

\begin{equation}
S(\omega)_{\text{modified}}[k] = 
\begin{cases} 
B(\omega)[k] & \text{if } \text{start\_index} \leq k < \text{end\_index} \\
S(\omega)[k] & \text{otherwise}
\end{cases} \label{eq:freq_replacement}
\end{equation}

\begin{enumerate}
    \setcounter{enumi}{5}
    \item \textbf{Reconstruction with IFFT:} The inverse FFT (IFFT) is applied to reconstruct the time-domain waveform:
\end{enumerate}

\begin{equation}
s_{\text{modified}}(t) = \text{IFFT}(S(\omega)_{\text{modified}}) \label{eq:ifft}
\end{equation}

\begin{figure*}[htbp]
  \centering
  \subfloat[Original fake sample]{%
    \includegraphics[width=0.32\linewidth]{Images/chap3/Spoof.png}
    \label{fig:original_spoof}
  }
  \hfill
  \subfloat[Frequency-domain artifact 2000--2500 Hz]{%
    \includegraphics[width=0.32\linewidth]{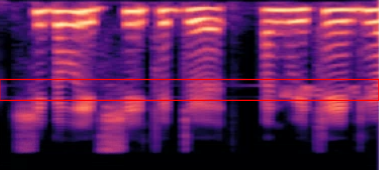}
    \label{fig:spoof_2000_2500}
  }
  \hfill
  \subfloat[Frequency-domain artifact 2000--3500 Hz]{%
    \includegraphics[width=0.32\linewidth]{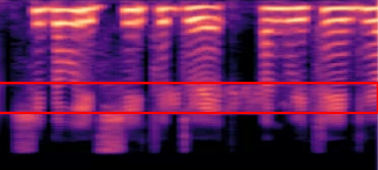}
    \label{fig:spoof_2000_3500}
  }

  \caption{Comparison of original and modified fake audio with frequency-domain artifacts. Horizontal distortions result from spectral modifications in the given frequency ranges, exposing inconsistencies in fake speech generation.}
  \label{fig:spoof_comparisons}
\end{figure*}

After this frequency swapping process, we extract Mel-spectrogram images as in the baseline method. We experiment with two frequency ranges: 2000-2500 Hz and 2000-3500 Hz. Fig. \ref{fig:spoof_comparisons} illustrates the generated horizontal artifact bands.

\subsubsection{Time Domain Artifact}

Unlike the frequency-domain artifacts, which create horizontal perturbations in the Mel-spectrogram images, time-domain artifacts introduce vertical patterns that reflect discontinuities and abrupt changes within the signal.

Time-domain artifacts are generated by segmentally replacing portions of a fake audio waveform with corresponding regions from a  real sample of the same speaker. This allows for controlled manipulation while preserving the overall speech structure. The process is described as follows:

\begin{enumerate}
    \item \textbf{Standardization:} Both  real and fake samples are standardized to 3 seconds in duration.
    \item \textbf{Window Selection:} A random segment is selected from the fake waveform for replacement.
    \item \textbf{Segment Replacement:} The selected segment in the fake sample is replaced with the corresponding segment from a  real sample of the same speaker ID.
    \item \textbf{Reconstruction:} The modified waveform is saved and subsequently converted into a Mel-spectrogram.
\end{enumerate}

Formally, let \( s(t) \) and \( b(t) \) represent the fake and  real audio waveforms in the time domain, respectively. The modification is performed as:

\begin{equation}
s_{\text{modified}}(t) =
\begin{cases} 
b(t) & \text{for } t \in [t_{\text{start}}, t_{\text{end}}] \\
s(t) & \text{otherwise}
\end{cases} \label{eq:time_swap}
\end{equation}

where \( t_{\text{start}} \) and \( t_{\text{end}} \) denote the randomly chosen segment boundary for replacement. The selection of speaker-matched  real and fake samples ensures that speaker identity remains constant, reinforcing identity independence. 

After generating the modified audio, we extract Mel-spectrogram images. This transformation allows the model to learn temporal inconsistencies as visually distinguishable vertical bands.

\begin{figure*}[htbp]
  \centering
  \subfloat[Original fake sample]{%
    \includegraphics[width=0.32\linewidth]{Images/chap3/Spoof.png}
    \label{fig:original_spoof_time}
  }
  \hfill
  \subfloat[Time-domain artifact 2000--2500 Hz]{%
    \includegraphics[width=0.32\linewidth]{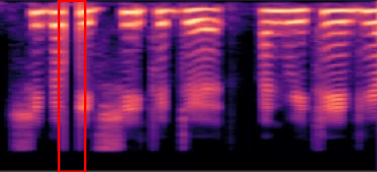}
    \label{fig:time_2000_2500}
  }
  \hfill
  \subfloat[Time-domain artifact 2000--3500 Hz]{%
    \includegraphics[width=0.32\linewidth]{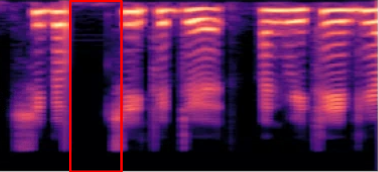}
    \label{fig:time_2000_3500}
  }

  \caption{Comparison of original and modified fake audio with time-domain artifacts. Vertical distortions are introduced by replacing audio segments in the given frequency ranges, creating structured disruptions that highlight synthetic inconsistencies.}
  \label{fig:time_artifacts}
\end{figure*}

Fig. \ref{fig:time_artifacts} illustrates the generated vertical artifact bands, highlighting the differences introduced by time-domain manipulations.

\subsubsection{Dynamic Frequency Swap Artifact}

To further enhance the variety of artifacts while preserving the speaker’s identity, we introduce  dynamic frequency swapping. Unlike the fixed-frequency band swap in the previous approach, this method selects a frequency range dynamically for each sample, ensuring that artifacts are diverse across different training instances. The key motivation behind this method is to create  varied yet controlled perturbations  that enhance the generalization of the Artifact Detection Module (ADM), preventing overfitting to specific artifact patterns. As with the fixed-frequency artifact, this technique  relies on randomly selecting a real sample from the same speaker ID as the fake sample.

\textit{Nyquist Frequency and Dynamic Band Selection:} The Nyquist frequency is a fundamental concept in signal processing, defined as half the sampling rate (\( \frac{SR}{2} \)). It represents the highest frequency that can be accurately represented in a sampled signal. Given that our audio data has a sampling rate of  16 kHz , the  Nyquist frequency is 8 kHz . 

To ensure meaningful modifications while maintaining speech intelligibility, we dynamically select a  start frequency in the range of 200 Hz to  70\% of the Nyquist frequency  (i.e., approximately  5.6 kHz ). The width of the swapped band is also chosen dynamically between 100 Hz and 500 Hz, ensuring variability across different samples. This approach introduces artifacts that are unpredictable but systematically constrained to realistic frequency regions. The process follows these key steps:
\begin{enumerate}
    \item \textbf{Standardization:} Both fake and  real samples are standardized to 3 seconds. 
    \item \textbf{Fourier Transform:} The Fast Fourier Transform (FFT) is applied to convert both waveforms to the frequency domain.
   \item \textbf{Dynamic Frequency Selection:} A random start frequency is selected from the range  200 Hz to 5.6 kHz. The band width is randomly chosen between  100 Hz and 500 Hz.
   \item \textbf{Component Replacement:} The identified frequency components in the fake sample are replaced with the corresponding components from the real sample of the same speaker ID.
   \item \textbf{Inverse FFT:} The modified signal is converted back to the time domain.
   \item \textbf{Normalization:} The waveform is normalized to prevent clipping.
   \item \textbf{Feature Extraction:} The resulting waveform is transformed into a Mel-spectrogram.
\end{enumerate}

Let \( s(t) \) and \( b(t) \) represent the  fake  and   real  waveforms, respectively. Their Fourier Transforms \( S(\omega) \) and \( B(\omega) \) are computed as:

\begin{equation}
S(\omega) = \text{FFT}(s(t)), \quad B(\omega) = \text{FFT}(b(t)) \label{eq:fft_dynamic}
\end{equation}

The Nyquist frequency is given by:

\begin{equation}
F_{\text{Nyquist}} = \frac{SR}{2}
\end{equation}
where \(SR\) is the sampling rate. 
Instead of a fixed swap, we dynamically select a start frequency and bandwidth from uniform distributions:

\begin{equation}
F_{\text{start}} \sim \mathcal{U}(200, 0.7 \cdot F_{\text{Nyquist}})
\end{equation}

\begin{equation}
\Delta F \sim \mathcal{U}(100, 500)
\end{equation}

Where \( \mathcal{U}(a, b) \) denotes a uniform distribution between \( a \) and \( b \). 
The start frequency is chosen between 200 Hz and 70\% of the Nyquist frequency (to avoid very low or very high frequencies). 
The bandwidth \( \Delta F \) is chosen between 100 Hz and 500 Hz to introduce a range of artifact sizes.

The  end frequency  is then calculated as:

\begin{equation}
F_{\text{end}} = \min(F_{\text{start}} + \Delta F, F_{\text{Nyquist}})
\end{equation}

This ensures that the selected band does not exceed the maximum representable frequency.

To locate the corresponding indices in the FFT representation, we compute the frequency bins as:

\begin{equation}
f_k = \frac{k \cdot SR}{N} \label{eq:freq_bins_dynamic}
\end{equation}

Where \( k \) is the index in the FFT array, \( SR \) is the sampling rate, \( N \) is the total number of samples.

We then determine the start and end indices of the frequency band:

\begin{equation}
\text{start\_index} = \min \left\{ k : f_k \geq F_{\text{start}} \right\} \label{eq:start_index_dynamic}
\end{equation}

\begin{equation}
\text{end\_index} = \min \left\{ k : f_k \geq F_{\text{end}} \right\} \label{eq:end_index_dynamic}
\end{equation}

Then, the frequency band in \( S(\omega) \) is replaced with the corresponding band from \( B(\omega) \), ensuring that only the randomly selected region is altered:

\begin{equation}
S(\omega)_{\text{modified}}[k] = 
\begin{cases} 
B(\omega)[k] & \text{for } \text{start\_index} \leq k < \text{end\_index} \\
S(\omega)[k] & \text{otherwise}
\end{cases} \label{eq:freq_replacement_dynamic}
\end{equation}

Finally, the modified frequency domain representation is transformed back into the time domain using Inverse FFT:

\begin{equation}
s_{\text{modified}}(t) = \text{IFFT}(S(\omega)_{\text{modified}}) \label{eq:ifft_dynamic}
\end{equation}

Following this transformation, the processed signal is converted into a Mel-spectrogram. The resulting artifact manifests as a horizontal perturbation at an unpredictable frequency location, ensuring diverse artifact placements across training samples. The Dynamic Frequency Swap allows for greater variability in artifact representation compared to fixed-band methods, providing diverse patterns across training instances. This variability strengthens the model's ability to learn forgery-related cues rather than memorizing dataset-specific features.

\begin{figure*}[htbp]
  \centering

  \begin{minipage}[b]{0.48\linewidth}
    \centering
    \subfloat[Original fake sample 1\label{fig:original_spoof_dynamic_1}]{
      \includegraphics[height=3cm]{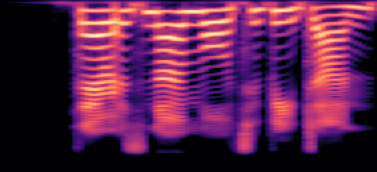}
    }
  \end{minipage}
  \hfill
  \begin{minipage}[b]{0.48\linewidth}
    \centering
    \subfloat[Dynamic Frequency Swap -- Modified sample 1\label{fig:dynamic_1}]{
      \includegraphics[height=3cm]{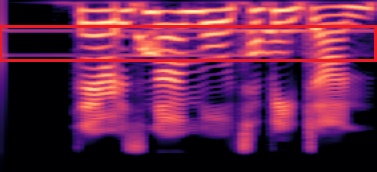}
    }
  \end{minipage}

  \vspace{5pt}

  \begin{minipage}[b]{0.48\linewidth}
    \centering
    \subfloat[Original fake sample 2\label{fig:original_spoof_dynamic_2}]{
      \includegraphics[height=3cm]{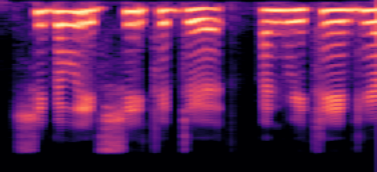}
    }
  \end{minipage}
  \hfill
  \begin{minipage}[b]{0.48\linewidth}
    \centering
    \subfloat[Dynamic Frequency Swap -- Modified sample 2\label{fig:dynamic_2}]{
      \includegraphics[height=3cm]{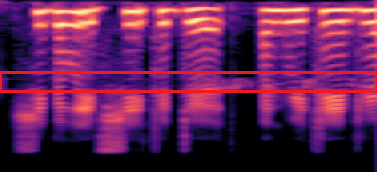}
    }
  \end{minipage}

  \caption{Comparison of original and dynamically modified fake samples using Dynamic Frequency Swap. This method injects artifacts by swapping frequency bands at randomized positions per sample, ensuring variation in horizontal artifact bands.}
  \label{fig:dynamic_freq_swap}
\end{figure*}

In addition to frequency and time domain manipulations, we also introduce an alternative augmentation method by blending real speech as background noise into fake samples. This simulates real-world acoustic conditions and tests the model’s resilience to background interference. To illustrate the impact of dynamic frequency swaps, Fig. \ref{fig:dynamic_freq_swap} presents a comparison between two original fake samples and their dynamically modified versions.

\subsubsection{Background Noise Injection}

To introduce additional variability in the artifacts while maintaining speaker identity consistency, we propose a  background noise augmentation  technique. This method involves adding a   real sample as background noise to a fake sample of the same speaker ID.

The primary objective of this approach is to  expose the model to a new type of perturbation similar to real recording noises while preserving the natural structure of the fake audio . This allows the Artifact Detection Module (ADM) to generalize better across unseen fake attacks. The process of generating this artifact follows these steps:

\begin{enumerate}
    \item \textbf{Standardization:} Both the fake and  real samples are padded or truncated to  3 seconds.
    \item \textbf{Noise Scaling:} The  real sample is  scaled down  by a noise level factor (\(\alpha\)) to ensure it acts as background noise.
    \item \textbf{Noise Addition:} The scaled  real audio is added to the fake sample.
    \item \textbf{Normalization:}  The final waveform is normalized to prevent clipping.
    \item \textbf{Feature Extraction:}  The processed waveform is converted into a Mel-spectrogram.
\end{enumerate}

Let \( s(t) \) represent the  fake  audio signal and \( b(t) \) represent the   real  audio signal from the same speaker ID. The background noise is scaled by a factor \( \alpha \), where \( 0 < \alpha < 1 \). The modified audio waveform is computed as:

\begin{equation}
s_{\text{modified}}(t) = s(t) + \alpha \cdot b(t) \label{eq:background_noise}
\end{equation}

Where \( \alpha \) is the noise scaling factor , empirically set to 0.2 to preserve speech intelligibility while introducing background perturbation. \( s(t) \) is the fake audio waveform. \( b(t) \) is the real background audio waveform.\( s_{\text{modified}}(t) \) is the final waveform with added background noise.

To ensure amplitude consistency, the modified waveform is  normalized  as follows:

\begin{equation}
s_{\text{normalized}}(t) = \frac{s_{\text{modified}}(t)}{\max(|s_{\text{modified}}(t)|)}
\end{equation}

This prevents signal clipping while preserving the noise perturbation.
 
Following noise addition, the processed signal is transformed into a Mel-spectrogram. The resulting  artifact manifests as a diffuse spectral perturbation, resembling natural background interference rather than structured spoofing artifacts.

\begin{figure*}[htbp]
  \centering

  \begin{minipage}[b]{0.48\linewidth}
    \centering
    \subfloat[Original fake sample 1\label{fig:original_spoof_background_1}]{
      \includegraphics[height=3cm]{Images/chap3/LA_T_1017386.png}
    }
  \end{minipage}
  \hfill
  \begin{minipage}[b]{0.48\linewidth}
    \centering
    \subfloat[Background Noise -- Modified sample 1 (Low)\label{fig:background_1}]{
      \includegraphics[height=3cm]{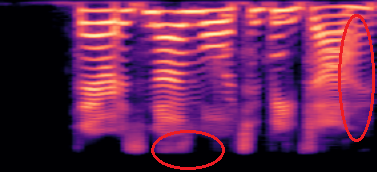}
    }
  \end{minipage}

  \vspace{5pt}

  \begin{minipage}[b]{0.48\linewidth}
    \centering
    \subfloat[Original fake sample 2\label{fig:original_spoof_background_2}]{
      \includegraphics[height=3cm]{Images/chap3/LA_T_1036415.png}
    }
  \end{minipage}
  \hfill
  \begin{minipage}[b]{0.48\linewidth}
    \centering
    \subfloat[Background Noise -- Modified sample 2 (High)\label{fig:background_2}]{
      \includegraphics[height=3cm]{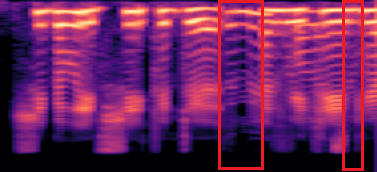}
    }
  \end{minipage}
  \caption{Comparison of original and modified fake audio with background noise artifacts. The presence of real noise introduces a diffused texture across the Mel-spectrogram, distinguishing it from synthetic modifications.}
  \label{fig:background_noise}
\end{figure*}

To illustrate the effect of background noise addition, Fig. \ref{fig:background_noise} presents a comparison between two original fake samples and their modified versions with added real background noise.

The subsequent section presents an evaluation of the proposed method in both intra-dataset and inter-dataset settings.

\section{Experimental Setup}
\label{sec:experimental}

To validate the proposed identity-independent framework, we conduct extensive experiments on four benchmark datasets: ASVspoof2019 (LA), ADD 2022, FoR, and In-The-Wild. Below, we describe the datasets, our preprocessing steps, and implementation details.

\subsection{Datasets}
\label{sec:datasets}

We use the ASVspoof2019 Logical Access (LA) subset \cite{nautsch2021asvspoof} for training and ADD 2022 \cite{yi2022add} for cross-dataset evaluation. ASVspoof2019 includes TTS and VC attacks, with metadata indicating speaker ID, spoofing method, and label (real or fake). ADD 2022 contains both real and fake samples with background noise and music.

To test generalizability and identity leakage, we also evaluate on Fake or Real (FoR) \cite{8906599} and In-The-Wild \cite{muller22_interspeech}. FoR contains diverse synthesis and conversion attacks, including unseen types. In-The-Wild offers real-world deepfakes with varied environments and transmission noise. Default splits are used, except for In-The-Wild, which we divide into 70/20/10 for training, validation, and testing. Dataset statistics are in Table \ref{table:datasets}.

\begin{table}[htbp]
\centering
\caption{Summary of dataset sizes (real vs. fake).}
\begin{tabular}{lcccccc}
\toprule
\multirow{2}{*}{Dataset} & \multicolumn{3}{c}{Real} & \multicolumn{3}{c}{Fake} \\
\cmidrule(lr){2-4}\cmidrule(lr){5-7}
 & Train & Val & Test & Train & Val & Test \\
\midrule
ASVspoof2019 LA & 2580 & 2548 & 7355 & 22800 & 22296 & 63882 \\
ADD 2022 & 3012 & 2307 & 300 & 24072 & 21295 & 700 \\
FoR & 26939 & 5400 & 2264 & 26927 & 5398 & 2370 \\
In-The-Wild & 13998 & 3945 & 2020 & 8247 & 2414 & 1155 \\
\bottomrule
\end{tabular}%

\label{table:datasets}
\end{table}

\subsection{Preprocessing}
\label{sec:preprocessing}

All audio is standardized to 3\,s, resampled to 16\,kHz, and converted to Mel-spectrogram images (magma colormap) sized for the CNN input (299\(\times\)299 for Xception). Dark purple indicates low intensity and bright yellow, high energy. See Fig.~\ref{fig:mel_spectrograms} for an example.

\subsection{Implementation Details} \label{sec:implementation}
We evaluate four CNN backbones on ASVspoof2019 as a preliminary test: Xception, EfficientNet-B3, ResNet50, and VGG16. Each is ImageNet‐initialized and trained for 50 epochs using BCE loss, batch normalization, dropout, L2 regularization, and a step-decay LR (\(10^{-3}\)). Table~\ref{tab:results} shows Xception yields the best F1 score(0.689), so we adopt it as our baseline. Table~\ref{tab:model_layers} details the architecture used for both the baseline and ADM enhanced models.

\begin{table} [!t]
  \centering
  \footnotesize
  \captionsetup{font=small}
  \caption{Summary of the Xception-Based Model Architecture}
  \label{tab:model_layers}
  \begin{tabularx}{\columnwidth}{Xc}
    \toprule
    \textbf{Component} & \textbf{Output Shape} \\
    \midrule
    Feature Extractor (Xception) & (None, 2048) \\
    Batch Normalization + Dropout & (None, 2048) \\
    Fully Connected (Dense 128) & (None, 128) \\
    Final Classification Layer & (None, 1) \\
    \midrule
    \textbf{Total Parameters} & 21.13M \\
    \textbf{Trainable} & 21.07M \\
    \textbf{Non-Trainable} & 58.88K \\
    \bottomrule
  \end{tabularx}
\end{table}

\begin{table}[!t]
\centering
\footnotesize
\caption{Backbone model comparison on ASVspoof2019. Accuracy and F1 scores are used to select the Xception architecture as the baseline for subsequent experiments.}
\label{tab:results}
\begin{tabularx}{\columnwidth}{Xccc}
\toprule
\textbf{Model} & \textbf{Accuracy} & \textbf{F1 Score} & \textbf{Params} \\
\midrule
Xception   & 0.9331 & \textbf{0.6892} & 22.9M \\
EfficientNet-B3 & 0.8960 & 0.0109 & 12.3M \\
ResNet50   & 0.8832 & 0.5491 & 25.6M \\
VGG16      & 0.8968 & 0      & 138M  \\
\bottomrule
\end{tabularx}
\end{table}

\subsubsection{Xception Baseline}

Xception \cite{chollet2017xception}, based on depthwise separable convolutions, separates spatial and channel correlations for efficient feature learning. Depthwise convolutions extract spatial features per channel, followed by pointwise (1×1) convolutions for cross-channel integration. This architecture improves computational efficiency and learning quality.

Like ResNet, Xception includes residual connections to enhance gradient flow and training convergence. These properties make it well-suited for extracting features from Mel-spectrograms in deepfake detection.

\subsubsection{Transfer Learning and Cross-dataset Results}

Based on its strong intra-dataset performance, the pre-trained Xception model, initially trained on the ASVspoof2019 dataset, is evaluated utilizing transfer learning for cross-dataset evaluations. Following \cite{dong2023implicit}, We fine-tune only the final dense layer on ADD 2022, FoR, and In-The-Wild, keeping the rest of the model frozen.

\textit{Baseline vs. ADM Training}: We first train Xception on real vs. fake. Then, freeze its feature extractor and train the ADM (same backbone) to classify “original fake vs. artifact-fake.” Finally, we unfreeze everything and re‐train on real-vs-fake classification using artifact-informed representations.


\section{Results and Discussion} \label{sec:results}

\subsection{Metric Analysis and Significance}
The evaluation metrics for our proposed method are F1 score, Precision, Recall, Equal Error Rate (EER), and Area Under the ROC Curve (AUC), each providing unique insights into the performance of deepfake detection models. F1 Score is particularly valuable in scenarios with class imbalance, common in deepfake detection tasks, as it harmonizes precision (minimizing false alarms) and recall (capturing maximum deepfakes).

EER reflects overall discriminative power independent of thresholds, identifying the point where false positive and negative rates coincide. Lower EER signifies robust general performance. AUC measures the model’s ability to rank samples correctly across all thresholds, with higher AUC demonstrating consistent classification performance.

\subsection{Comparative Analysis Across Datasets}
The variance in results across ASVspoof2019, ADD 2022, FoR, and In-The-Wild datasets highlights how different data characteristics influence metric performance.

During intra-dataset evaluations, all models, particularly the Baseline and Dynamic Swap, performed robustly, with high F1 scores, precision, recall, and notably low EER values (around 0.1), coupled with AUCs exceeding 0.94. The high performance is expected since models were optimized on ASVspoof2019, reflecting effective learning of intra-domain artifacts. Intra-dataset results are reported in Table~\ref{tab:results_asvspoof}, serving as a baseline before applying ADM-based enhancements.

\begin{table}[htbp]
\centering
\caption{Intra-Dataset Evaluation on ASVspoof2019}
\begin{tabular}{lccccc}
\toprule
\textbf{Model} & \textbf{F1} & \textbf{Precision} & \textbf{Recall} & \textbf{EER} & \textbf{AUC} \\
\midrule
Baseline & 0.6805 & 0.5812 & 0.8205 & 0.1224 & 0.9409 \\
Horizontal 2000–2500 & 0.7060 & 0.6071 & 0.8435 & 0.1003 & 0.9622 \\
Horizontal 2000–3500 & 0.6923 & 0.5736 & 0.8729 & 0.0980 & 0.9586 \\
Vertical 2000–2500 & 0.7335 & 0.6885 & 0.7849 & 0.1072 & 0.9617 \\
Vertical 2000–3500 & 0.6749 & 0.5697 & 0.8277 & 0.1212 & 0.9471 \\
Background Noise & 0.6286 & 0.4889 & 0.8801 & 0.1150 & 0.9510 \\
Dynamic Swap & 0.7416 & 0.6997 & 0.7890 & 0.1154 & 0.9543 \\
\bottomrule
\end{tabular}%

\label{tab:results_asvspoof}
\end{table}

\begin{table*}[htbp]
\centering
\caption{Cross-Dataset Evaluation Results on ADD 2022, FoR, and In-The-Wild}
\resizebox{\textwidth}{!}{%
\begin{tabular}{lccccccccccccccc}
\toprule
\textbf{Model} 
& \multicolumn{5}{c}{\textbf{ADD 2022}} 
& \multicolumn{5}{c}{\textbf{FoR}} 
& \multicolumn{5}{c}{\textbf{In-The-Wild}} \\
\cmidrule(lr){2-6} \cmidrule(lr){7-11} \cmidrule(lr){12-16}
& \textbf{F1} & \textbf{Prec.} & \textbf{Rec.} & \textbf{EER} & \textbf{AUC}
& \textbf{F1} & \textbf{Prec.} & \textbf{Rec.} & \textbf{EER} & \textbf{AUC}
& \textbf{F1} & \textbf{Prec.} & \textbf{Rec.} & \textbf{EER} & \textbf{AUC} \\
\midrule
Baseline 
& 0.1726 & 0.8056 & 0.0967 & 0.4300 & 0.6008 
& 0.5854 & 0.6466 & 0.5341 & \textbf{0.4927} & 0.5095 
& 0.7828 & 0.8129 & 0.7540 & 0.3767 & 0.6701 \\

Horizontal 2000–2500 
& 0.0000 & 0.0000 & 0.0000 & 0.4948 & 0.4908 
& 0.5640 & 0.6514 & 0.4991 & 0.5261 & 0.4652 
& 0.7602 & 0.8135 & 0.7125 & 0.3663 & 0.6994 \\

Horizontal 2000–3500 
& 0.2123 & 0.6552 & 0.1267 & 0.3538 & 0.6921 
& 0.5836 & 0.6094 & 0.5601 & 0.5412 & 0.4281 
& \textbf{0.8244} & \textbf{0.8605} & \textbf{0.7913} & \textbf{0.2635} & \textbf{0.8171} \\

Vertical 2000–2500 
& 0.0392 & \textbf{1.0000} & 0.0200 & 0.4621 & 0.5643 
& 0.5119 & 0.6212 & 0.4387 & 0.5555 & 0.4224 
& 0.7919 & 0.8377 & 0.7516 & 0.3487 & 0.7053 \\

Vertical 2000–3500 
& 0.1818 & 0.7561 & 0.1033 & 0.3507 & 0.7210 
& 0.4616 & 0.6114 & 0.3665 & 0.4924 & \textbf{0.5100} 
& 0.8019 & 0.8519 & 0.7582 & 0.3183 & 0.7364 \\

Background Noise 
& 0.3540 & 0.5704 & \textbf{0.2567} & 0.3424 & 0.6921 
& 0.5035 & 0.5876 & 0.4418 & 0.5583 & 0.4256 
& 0.7758 & 0.8004 & 0.7532 & 0.4504 & 0.5777 \\

Dynamic Swap 
& \textbf{0.2305} & 0.8511 & 0.1333 & \textbf{0.2700} & \textbf{0.8078} 
& \textbf{0.6042} & \textbf{0.6651} & \textbf{0.5536} & 0.5915 & 0.4035 
& 0.8132 & 0.8537 & 0.7773 & 0.3307 & 0.7123 \\
\bottomrule
\end{tabular}%
}
\label{tab:results_cross_full}
\end{table*}

The substantial degradation of the performance on ADD 2022 (F1 score dropping to 0.1726 and high EER at 0.4300) highlights significant cross-domain challenges. In contrast, the Dynamic Swap model showed a marked improvement (F1 score of 0.2305, EER of 0.2700, and notably high AUC of 0.8078). This indicates that dynamic artifact generation helps reduce speaker identity bias, improving generalization. The reason for metric variance, such as lower F1 score but higher AUC and lower EER, is due to the model's sensitivity to dataset-specific threshold calibration, emphasizing the advantage of EER and AUC in assessing potential model performance in cross-dataset scenarios. Due to FoR’s lack of consistent and learnable artifact patterns, the lowest cross-dataset F1 scores across all models, including the baseline (F1 = 0.5854, EER = 0.4927) is observed. The Horizontal 2000-3500 model achieved the highest F1 score (0.8244), significantly better than baseline when evaluated on In-The-Wild. Its remarkably low EER (0.2635) and high AUC (0.8171) underscore its capability to effectively generalize artifact detection.

\subsection{Visual Analysis using t-SNE}
\label{sec:tsne}

\begin{figure*}[htbp]
  \centering
  \subfloat[Baseline - ADD 2022]{
    \includegraphics[width=0.31\textwidth]{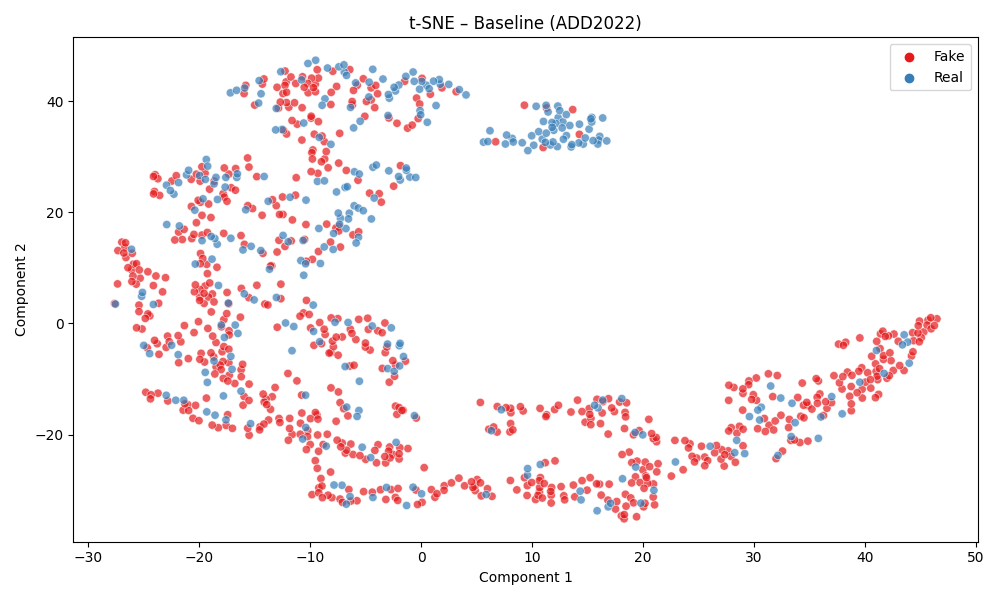}
  }\hfill
  \subfloat[Baseline - FoR]{
    \includegraphics[width=0.31\textwidth]{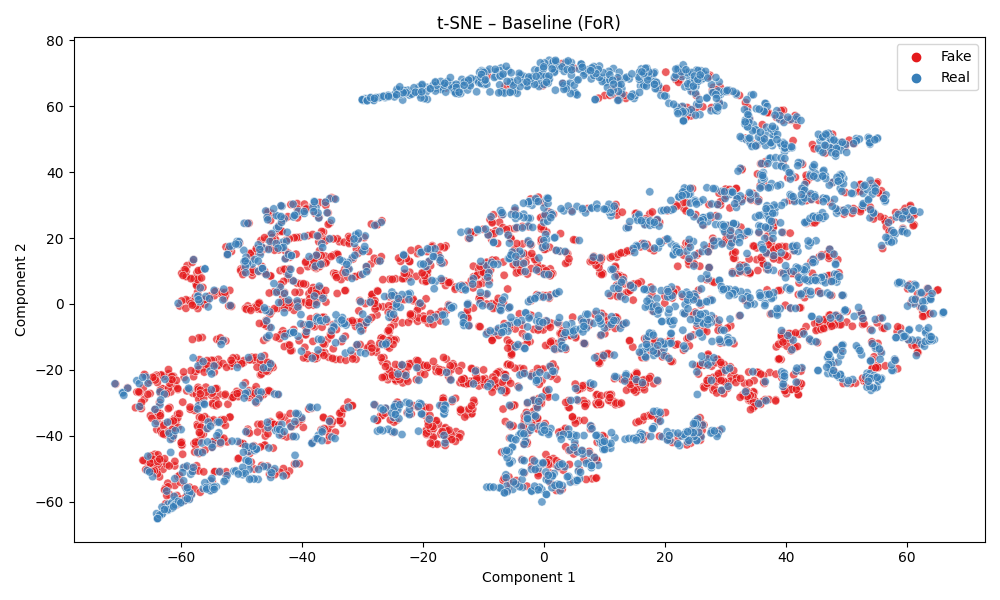}
  }\hfill
  \subfloat[Baseline - In‑The‑Wild]{
    \includegraphics[width=0.31\textwidth]{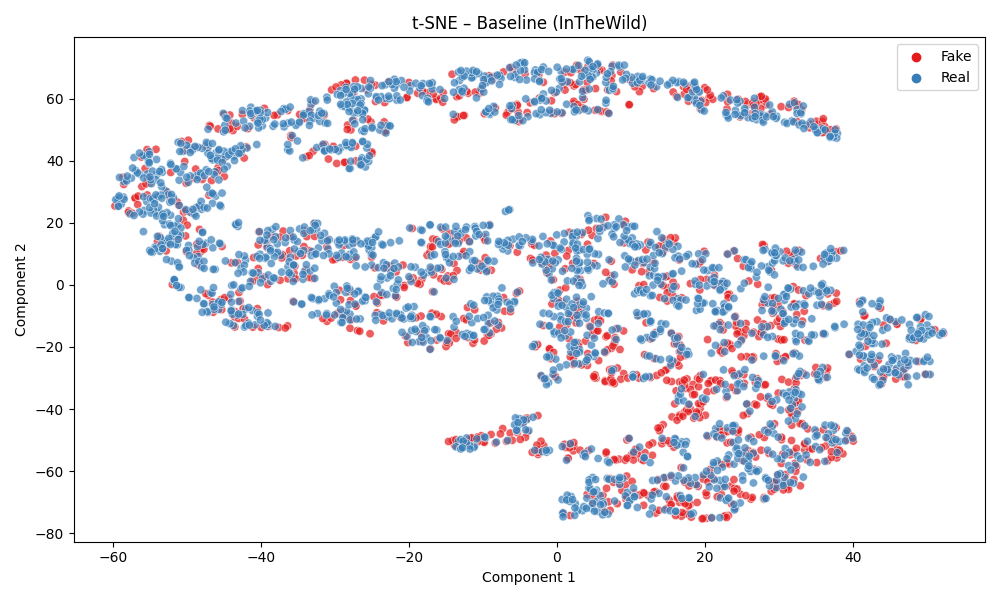}
  }
  \vspace{0.8em}

  \subfloat[Dynamic Swap - ADD 2022]{
    \includegraphics[width=0.31\textwidth]{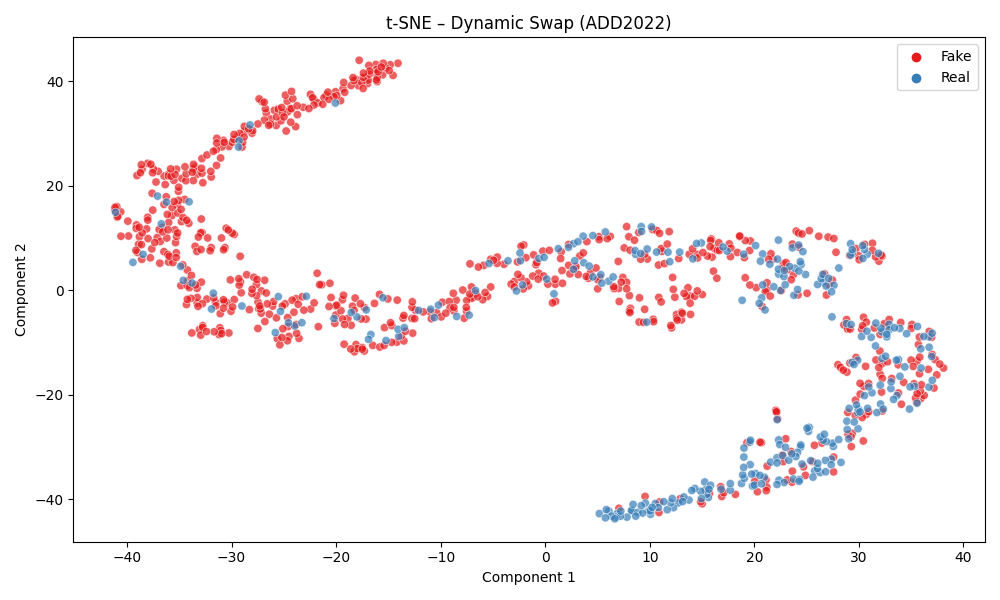}
  }\hfill
  \subfloat[Dynamic Swap - FoR]{
    \includegraphics[width=0.31\textwidth]{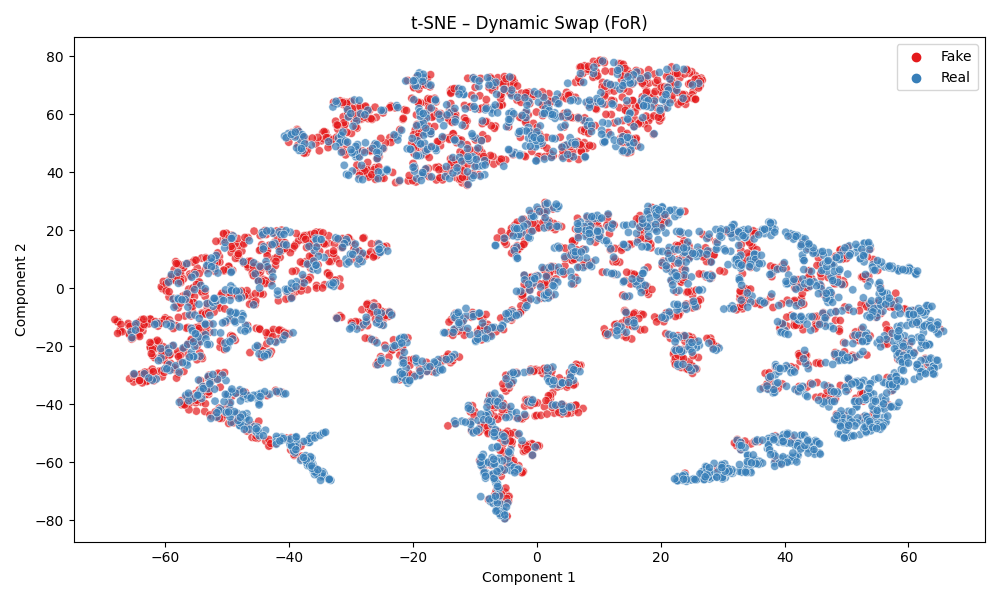}
  }\hfill
  \subfloat[Horizontal 2000–3500 Hz\\In‑The‑Wild]{
    \includegraphics[width=0.31\textwidth]{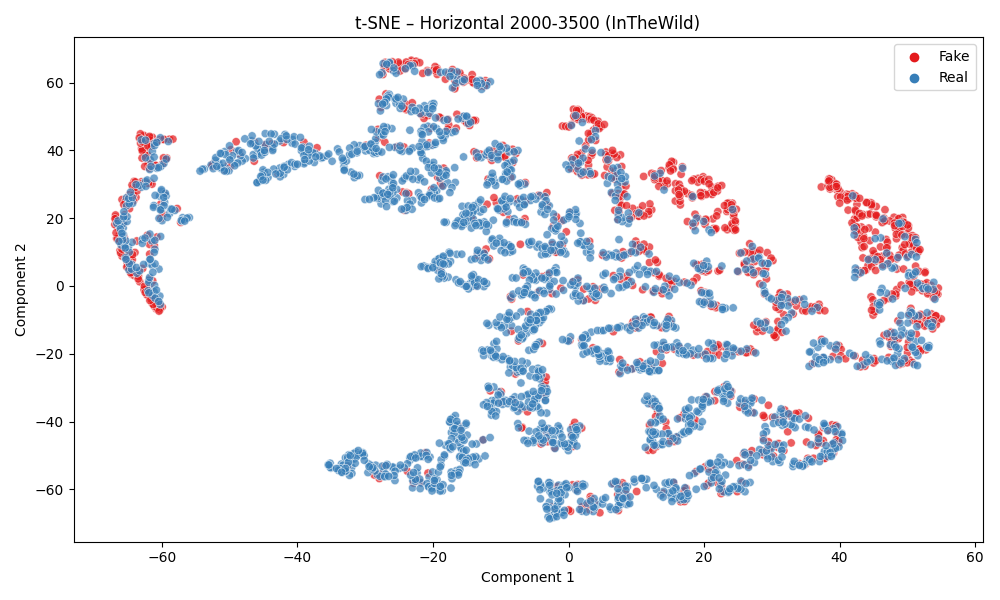}
  }
  \caption{t‑SNE projections of the last layer embeddings for cross dataset evaluations. Top row shows baseline model trained only on ASVspoof2019 while the bottom row displays the best ADM‑enhanced model for each dataset (Dynamic Swap for ADD 2022/FoR; Horizontal 2000–3500 Hz for In‑The‑Wild).  
  Red = fake, Blue = real. ADM training yields markedly better class separation, visually confirming the quantitative gains in F1 and EER as seen in the results.}
  \label{fig:tsne_grid}
\end{figure*}

To complement the numerical metrics, we visualize the embeddings in the last layer of each model with t‑SNE. Figure \ref{fig:tsne_grid} shows the baseline features on the three cross‑dataset test sets along with the best ADM‑enhanced configuration for each of them (Dynamic Swap with ADD 2022 and FoR, Horizontal 2000–3500 Hz with In‑The‑Wild).

Across all corpora the baseline clusters real and fake points in overlapping manifolds, indicating that the network relies on speaker identity cues rather than manipulation artifacts. This can be seen clearly in clusters of In‑The‑Wild where red and blue points are almost inseparable, mirroring the relatively high EER of 0.377 in Table \ref{tab:results_cross_full}. After artifact‑centered fine‑tuning, separation improves noticeably. For the ADD 2022 dataset, Dynamic Swap pushes most fake embeddings into the distinct side of the manifold, consistent with the EER reduction (0.43 $\rightarrow$ 0.27). As mentioned previously, FoR is a challenging corpus. Dynamic Swap reduces overlap and raises F1 score from 0.5854 to 0.6042 while reducing the EER, confirming that artifact cues improve discrimination. Horizontal 2000-3500 Hz yields two well‑defined lobes with minimal mixing in case of In-The-Wild dataset, consistent with highest cross‑dataset F1 (0.824) and lowest EER (0.2635).

These visualizations confirm that steering the network toward artifact regions suppresses implicit identity leakage and yields decision boundaries that generalize beyond the training domain.

\subsection{Justification of Metric Variations and Generalization}
The differential performance of metrics across datasets underscores the necessity of using multiple evaluation approaches:

EER and AUC become particularly important when dealing with cross-dataset generalization, as they indicate how robustly a model differentiates between classes without relying heavily on threshold calibration. For instance, the Dynamic Swap model's superior AUC and EER on ADD 2022 highlight its improved inherent discriminative capabilities, making it more robust for real-world applications where thresholds may vary.

F1 score, however, provides immediate operational insight at fixed thresholds, valuable when practical deployment constraints limit extensive threshold optimization. Its effectiveness in the In-The-Wild dataset signifies real-world applicability, showcasing a balance between precision and recall that can be directly leveraged for immediate deployment. The analysis demonstrates clear trade-offs and strengths across evaluation metrics, emphasizing their complementary roles in understanding model effectiveness. Artifact-based strategies, particularly Dynamic Swap and Horizontal modifications, consistently enhanced generalization performance across diverse testing scenarios.These results support our central hypothesis: that implicit speaker identity leakage undermines generalization, and identity-independent artifact learning improves robustness across datasets. While performance varied across datasets such as ADD 2022, FoR, and In-The-Wild, the consistent benefits of our artifact-based augmentation highlight its potential to develop more speaker-agnostic and transferable detection models.

\section{Conclusion and Future Work}
\label{sec:conclusion}

This paper introduced an identity-independent deepfake audio detection framework that directly tackles the challenge of \textit{identity leakage}, the model's tendency to rely on speaker-specific cues instead of manipulation-related features. By incorporating Artifact Detection Modules (ADMs) and a range of \textit{artifact generation} strategies (frequency swaps, time swaps, dynamic frequency swaps, and background noise), the proposed approach encourages learning that prioritizes tampering evidence over identity traits. Comprehensive experiments on ASVspoof2019, ADD 2022, FoR, and In-The-Wild demonstrate consistent improvements in cross-dataset generalization. Notably, the Dynamic Frequency Swap strategy achieves the lowest EER and highest AUC in most scenarios, highlighting its adaptability to unseen synthesis methods. In future, we aim to enhance generalization by refining training objectives that better capture subtle spoofing cues across domains. We also plan to investigate additional artifact types that reflect emerging manipulation trends.


\end{document}